\documentclass[a4paper]{jpconf}
\usepackage{graphicx}
\usepackage{multicol}
\usepackage{amsthm}
\usepackage{amsmath}
\usepackage{amssymb}
\usepackage{tabularx,slashed}
\usepackage{url}
\newcommand{\bea}{\begin{eqnarray}}
\newcommand{\eea}{\end{eqnarray}}

\newcommand{\nn}{\nonumber}
\begin{document}
\title{
\begin{flushright}
\ \\*[-80pt] 
\begin{minipage}{0.2\linewidth}
\normalsize
%arXiv:YYMM.NNNN \\
HUPD-2212 \\*[5pt]
\end{minipage}
\end{flushright}
Effective theory for universal seesaw model and FCNC}
\author{Takuya Morozumi $^1$$^2$,
 Apriadi Salim Adam$^3$, \\  Yuta Kawamura$^4$,
Albertus Hariwangsa Panuluh$^1$$^5$,\\
Yusuke Shimizu$^1$$^2$ and
 Kei Yamamoto$^6$
}
\address{$^1$ Physics Program, Graduate School of Advanced Science and Engineering,
Hiroshima University, Higashi-Hiroshima 739-8526, Japan}
\address{$^2$ Core of Research for the Energetic Universe,
Hiroshima University, Higashi-Hiroshima 739-8526, Japan}
\address{$^3$ Research Center for Quantum Physics, National Research and Innovation Agency (BRIN) 
, South Tangerang 15314, Indonesia}
\address{$^4$ Yokkaichi,  Mie Prefecture, Japan}
\address{$^5$ Department of Physics Education, Sanata Dharma University, Paingan, Maguwohardjo, Sleman, Yogyakarta 55282, Indonesia}
\address{$^6$ Department of Global Environment Studies, Hiroshima Institute of Technology, Hiroshima 731-
5193, Japan }
\ead{morozumi@hiroshima-u.ac.jp, apriadi.salim.adam@brin.go.id, kawamura1994phy@gmail.com, panuluh-albertus@hiroshima-u.ac.jp,
yu-shimizu@hiroshima-u.ac.jp, k.yamamoto.e3@cc.it-hiroshima.ac.jp
}
\begin{abstract}
We study the quark sector of the universal seesaw model  with SU(2)$_L$ $\times$ SU(2)$_R$ $\times$ $U(1)$.
The model incorporates the seesaw mechanism with the vector-like quarks (VLQs). The purpose of this work is
to study the model with the effective theory.  After integrating the heavy five VLQs, we derive the 
effective theory with four up-type quark and three down type quark. In this work, the
FCNC of Z boson for top quark and top$^\prime$ quark is derived.
\end{abstract}

\section{Introduction}
Though the standard model is a very successful theory, 
the origin of the flavor and mass hierarchy of quarks can be explained only  by tuning the Yukawa coupling
\cite{kobayashi}.
For instance up quark mass and top quark mass are respectively given as,
\bea
&& m_u=y_u \frac{v}{\sqrt{2}} < m_t=y_t \frac{v}{\sqrt{2}} \nn \\
&& y_u \simeq  1.25 \times 10^{-5} y_t , \quad y_t \simeq 0.99,
\eea
where we use the top quark mass from the direct measurement $m_t=172.69$(GeV) and
the up quark mass from the $\overline{MS}$ scheme $m_u(2 {\rm GeV})=2.16$(MeV)
\cite{ParticleDataGroup:2022pth}.
We also use $v=246.22$(GeV) to derive Yukawa coupling of the top quark.
The universal seesaw model explains the smallness of the mass of the up quark with a tiny ratio
of SU(2)$_R$ breaking scale and a SU(2) singlet vector-like quark (VLQ) mass $M_U$.
The standard model Yukawa coupling $y_u$ is given by the seesaw like formula,
\bea
&&y_u=y_{uL} \left(\frac{v_R}{\sqrt{2} M_U} \right) y_{uR}\simeq  \left(\frac{v_R}{\sqrt{2}M_U} \right)\simeq 10^{-5}, \nn \\
&& y_{uL} \simeq y_{uR}\simeq O(1),
\eea 
where $y_{uL}$ and $y_{uR}$ are mixing type Yukawa couplings between the ordinary quark and  VLQ.
We study the quark sector of the universal seesaw model  with SU(2)$_L$ $\times$ SU(2)$_R$ $\times$
U(1) \cite{ber, raj,chang,dav}.  Our aim is to construct the effective theory obtained after integrating  VLQs with their masses  larger than SU(2)$_R$ breaking scale.
Then, we study the flavor structure of the effective theory.
The same model was investigated with the full theory \cite{
Koide:1995pb,
Morozumi:1997af, Kiyo:1998zm}.
%These guidelines show how to prepare articles for publication in \jpcs\ using \LaTeX\ so they can be published quickly and accurately. Articles will be %refereed by the \corg s but the accepted PDF will be published with no editing, proofreading or changes to layout. It is, therefore, the author's %responsibility to ensure that the content and layout are correct.  This document has been prepared using \cls\ so serves as a sample document. The %class file and accompanying documentation are available from \verb"http://jpcs.iop.org".
\section{The Lagrangian of the model }
We first present the quark and gauge sector. 
The two SU(2) doublet Higgs fields are introduced. 
$\phi_L$ stands for SU(2)$_L$ doublet and $\phi_R$ for SU(2)$_R$.
Their vacuum expectation values (vevs) $v_L$ and $v_R$ repectively break SU(2)$_L$ and SU(2)$_R$.
From the Higgs potential study, one can show that the vevs
satisfy  $v_R \gg v_L$.   
The ordinary quarks $\psi_L$($\psi_R$) are also  SU(2)$_L$
(SU(2)$_R$) doublets.   The six  VLQs ($U_1\sim U_3 , D_1 \sim D_3$) 
are also introduced and the Lagrangian is, 
\bea
{\cal L}_{Doublets}&=& 
\sum_{i=1}^3 \overline{\psi_{Li}} \left( i \slashed{\partial}-g_L \slashed{W}_L 
-g_1 \frac{1}{6} \slashed{B}_1\right)  \psi_{Li} + ( L \rightarrow R), \nn \\
 {\cal L}_{VLQ}&=&\sum_{I=1}^3 \overline{U_I} \left( i \slashed{\partial}-g_1  \slashed{B}_1\frac{2}{3}-M_{U_I}\right)  U_I 
+\sum_{I=1}^3 \overline{D_I} \left( i \slashed{\partial}+g_1  \slashed{B}_1\frac{1}{3}-M_{D_I}\right)  D_I , \nn \\
{\cal L}_{VLQ-Doublets}&=&-y^u_{L i J} \overline{\psi_{i L}} \tilde{\phi}_L U_{J R}-y^u_{R iJ} \overline{\psi_{i R}} \tilde{\phi}_R U_{J L}-h.c.
\nn \\
&&-y^d_{L i J} \overline{\psi_{i L}} \phi_L D_{J R}-y^d_{R i J} \overline{\psi_{i R}} \phi_R D_{J L}-h.c. ,
\eea
 where $g_L$, $g_R$ and $g_1$ denote the gauge couplings for SU(2)$_L$, SU(2)$_R$ and $U(1)$, respectively.
\section{The mixing of the Neutral Gauge Bosons}
The symmetry breaking of  SU(2)$_L \times$ SU(2)$_R \times$ U(1) into U(1)$_{em}$ leads to the following relation between weak eigenstates $(W^3_L, B_1. W^3_R)$ and mass eigenstates $(Z, A, Z^\prime)$ of the three neutral gauge bosons,
\bea
&&
\begin{pmatrix}
W_{L\mu}^3 \\
B_{1 \mu}  \\
W_{R\mu}^3 
\end{pmatrix}=
O_{23}(\theta_{W_R}) O_{12}(-\theta_W) O_{13}(\theta_{13}) 
\begin{pmatrix}
Z_\mu \\
A_{\mu}  \\
Z^\prime_\mu  
\end{pmatrix} ,
\eea
where the rotation matrices are $O_{23}(\theta) =\begin{pmatrix}1 & 0 & 0 \\
 0 & \cos \theta  & -\sin \theta \\
0 &\sin \theta & \cos \theta \end{pmatrix},
O_{12}(\theta) =\begin{pmatrix}\cos \theta  &  -\sin \theta & 0 \\
 \sin \theta &  \cos \theta  & 0 \\
0 & 0 & 1 \end{pmatrix} $, $
O_{13}(\theta) =\begin{pmatrix}\cos \theta  & 0 & -\sin \theta \\
 0 & 1  & 0 \\
 \sin \theta & 0 & \cos \theta \end{pmatrix}$ and 
the mixing angles are given as
$\tan \theta_{W_R}=\frac{g_1}{g_R},  \tan \theta_{W}=\frac{g^\prime}{g_L}, 
\tan 2 \theta_{13} = \frac{\sin^2 \theta_{W_R} \sin 2\theta_{W_R}}{\sin \theta_W}\frac{v_L^2}{v_R^2}$.
We note that $\theta_{13}$ is suppressed by the ratio of vevs $\frac{v_L^2}{v_R^2}$.
With the mixing angles, the standard model-like U(1)$_{Y}$ hypercharge and the electromagnetic charge are given by 
$
g^\prime=g_1 \cos \theta_{W_R},   e=g^\prime \cos \theta_{W}
$.
The isospin current of Z boson is, 
\bea
{\cal L}_{{\rm Z}  I_3}&=& -  \left\lbrace  
\frac{1}{2 \cos \theta_W}\left(  \overline{u^i_L} \gamma_\mu u^i_L-\overline{d^i_L} \gamma_\mu d^i_L\right)   \left( g_L \cos \theta_{13} + e \tan\theta_{W_R} \sin \theta_{13}\right)  \right. \nn \\
&&\left.+  \frac{g_R}{2 \cos \theta_{W_R}} \left(  \overline{u^i_R} \gamma_\mu u^i_R-\overline{d^i_R} \gamma_\mu d^i_R\right)  \sin \theta_{13} \right\rbrace Z^\mu.
\label{eq:I3}
\eea
\section{Effective Lagrangian}
We integrate the five VLQs except a up-type VLQ denoted by $U_3$ which mass parameter $M_{U3}$  is smaller than $v_R$.
The following effective Lagrangian is obtained,
\bea
 {\cal L}_{\rm eff}&=&
\sum_{i=1}^3 \overline{\psi_{Li}} \left( i \slashed{\partial}-g_L \slashed{W}_L 
-g_1 \frac{1}{6} \slashed{B}_1\right)  \psi_{Li} + ( L \rightarrow R) \nn \\
&-&\sum_{ij} \frac{v_L}{\sqrt{2}} \overline{u_{L i}}u_{R j} y^{u}_{ij} -h.c.
-\sum_{ij} \frac{v_L}{\sqrt{2}} \overline{d_{L i}}d_{R j} y^{d}_{ij} -h.c. \nn \\
&-& \sum_{i} \frac{y^u_{L i3} v_L}{\sqrt{2}} \overline{u_{Li}} U_{R3}
- \sum_{i} \frac{y^u_{R i3} v_R}{\sqrt{2}}  \overline{u_{R i }} U_{L 3}-h.c.-\overline{U}_3 M_{U_3} U_3,
\label{eq:Leff}
\eea
where we have ignored the terms suppressed by a factor $\frac{v_L^2}{M_X^2} \ll  \frac{v_R^2}{M_X^2} \ll 1$ and standard model like Yukawa couplings for light quarks $u,d,c,s,b$ are given by,
\bea
y^u_{ij}&=&-\sum_{\alpha=1}^{2} y^u_{L i \alpha} y^{u \ast}_{R j \alpha} \frac{v_R}{\sqrt{2}M_{U \alpha}}, \quad
y^d_{ij}=-\sum_{\alpha=1}^{3} y^d_{L i \alpha} y^{d \ast}_{R j \alpha} \frac{v_R}{\sqrt{2}M_{D \alpha}},
\eea
where we have substituted the vevs to two Higgs doublets. 
${\cal L}_{\rm eff}$ includes four up-type quarks and  their $4 \times 4$ mass matrix is given by,
\bea
&& {\cal L}_{\rm eff\ up-type\ mass}=
-\begin{pmatrix} \overline{u_{Li}} & \overline{U_{L3}} \end{pmatrix}  M_{{\cal U}} 
\begin{pmatrix}
u_{j R} \\
{U}_{R3} 
\end{pmatrix}
, 
 M_{{\cal U}} =
\begin{pmatrix}  -\sum_{\alpha=1}^{2} \frac{ {\bf y}^u_{L  \alpha} {\bf y}^{u \dagger}_{R  \alpha}  v_L v_R}{{2}M_{U \alpha}} &  \frac{{\bf y}^u_{L3} v_L}{\sqrt{2}}  \\                     
\frac{{\bf y}^{u \dagger}_{R3} v_R}{\sqrt{2}} & M_{U_3}
\end{pmatrix}, \eea
where 
$
\frac{{\bf y}^{uT}_{L(R)3} v_L}{\sqrt{2}}=\begin{pmatrix} \frac{{y}^{u }_{L(R) 13} v_L}{\sqrt{2}} &  \frac{{y}^{u}_{L(R)23} v_L}{\sqrt{2}} & \frac{{ y}^{u}_{L(R) 33} v_L}{\sqrt{2}}\end{pmatrix}
$.
We apply the following bi-unitary transformation on 
$ {M}_{\cal U} $ with two $ 3 \times 3$ unitary matrices
$V^u$ and $W^u$,
\bea
&& \begin{pmatrix} V^{u \dagger} & 0_{3\times 1} \\ 0_{1 \times 3} & 1 \end{pmatrix} 
 {M}_{\cal U}\begin{pmatrix} W^{u} & 0_{3\times 1} \\ 0_{1 \times 3} & 1 \end{pmatrix}
=\begin{pmatrix}  -V^{u \dagger}(\sum_{\alpha=1}^{2} \frac{ {\bf y}^u_{L  \alpha} {\bf y}^{u \dagger}_{R  \alpha}  v_L v_R}{{2}M_{U \alpha}})  W^{u}  &  V^{u \dagger}\frac{{\bf y}^u_{L3} v_L}{\sqrt{2}}  \\                     
\frac{{\bf y}^{u \dagger}_{R3} v_R}{\sqrt{2}} W^{u}& M_{U_3}  \end{pmatrix} \nn \\
&&= \begin{pmatrix}  0  & 0 & 0  & 0   \\        
 0 & 0 & 0 & 0 \\
0 & 0 & 0 &   \frac{|{\bf y}^u_{L3}|v_L}{\sqrt{2}}    \\     
 0 & 0 & \frac{|{\bf y}^u_{R3}|v_R}{\sqrt{2}}  & M_{U_3}
\end{pmatrix} +
\begin{pmatrix}  -V^{u \dagger}(\sum_{\alpha=1}^{2} \frac{ {\bf y}^u_{L  \alpha} {\bf y}^{u \dagger}_{R  \alpha}  v_L v_R}{{2}M_{U \alpha}})  W^{u}  &  0_{3 \times 1}  \\                     
  0_{1 \times 3}  & 0
\end{pmatrix}.
\label{eq:Mu}
\eea
Below we diagonalize the first term of Eq.(\ref{eq:Mu}) to obtain the spectrum of the heavier up-type quark. The second term is ignored since the contribution is suppressed by $\frac{1}{M_{U \alpha}}$ $(\alpha=1,2)$. 
With the bi-orthogonal transformation, it is diagonalized as,
\bea
&&
 \begin{pmatrix}  
1 & 0 & 0  & 0   \\        
 0 & 1 & 0 & 0 \\
0 & 0 & \cos \theta_R & - \sin \theta_R \\     
 0 & 0 &\sin \theta_R  &\cos \theta_R
 \end{pmatrix} \begin{pmatrix} 
 1  & 0 & 0  & 0   \\        
 0 & 1 & 0 & 0 \\
0 & 0 & \cos \theta_l &  \sin \theta_l \\     
 0 & 0 &-\sin \theta_l &\cos \theta_l
\end{pmatrix} \nn \\
&& \times
\begin{pmatrix}  0  & 0 & 0  & 0   \\        
 0 & 0 & 0 & 0 \\
0 & 0 & 0 &   \frac{|{\bf y}^u_{L3}|v_L}{\sqrt{2}}    \\     
 0 & 0 & \frac{|{\bf y}^u_{R3}|v_R}{\sqrt{2}}  & M_{U_3}
\end{pmatrix} 
\begin{pmatrix} 
 1 & 0 & 0  & 0   \\        
 0 & 1 & 0 & 0 \\
0 & 0 & 0 &  1 \\     
 0 & 0 &1  & 0
\end{pmatrix}
\begin{pmatrix} 
 1  & 0 & 0  & 0   \\        
 0 & 1 & 0 & 0 \\
0 & 0 & \cos \theta_R &  \sin \theta_R \\     
 0 & 0 &-\sin \theta_R  & \cos \theta_R
\end{pmatrix} \nn \\
&=& 
 \begin{pmatrix}  
1 & 0 & 0  & 0   \\        
 0 & 1 & 0 & 0 \\
0 & 0 & \cos \theta_L & - \sin \theta_L \\     
 0 & 0 &\sin \theta_L  &\cos \theta_L
\end{pmatrix}
\begin{pmatrix}  0  & 0 & 0  & 0   \\        
 0 & 0 & 0 & 0 \\
0 & 0 &  m_L  &  0  \\     
 0 & 0 & M_{U_3} & m_R 
\end{pmatrix} 
\begin{pmatrix} 
 1  & 0 & 0  & 0   \\        
 0 & 1 & 0 & 0 \\
0 & 0 & \cos \theta_R &  \sin \theta_R \\     
 0 & 0 &-\sin \theta_R  & \cos \theta_R
\end{pmatrix}
=
\begin{pmatrix} 
 0  & 0 & 0  & 0   \\        
 0 & 0 & 0 & 0 \\
0 & 0 & m_t  & 0 \\     
 0 & 0 &0 & m_{t^\prime}
\end{pmatrix}, \nn \\
&& m_{t(t^\prime)}=\frac{\sqrt{M_{U_3}^2+(m_R+m_L)^2}}{2}\mp \frac{\sqrt{M_{U_3}^2+(m_R-m_L)^2}}{2},
\eea
where
$
m_{L}= \frac{|{\bf y}^u_{L3}|v_L}{\sqrt{2}} ,  m_{R}= \frac{|{\bf y}^u_{R3}|v_R}{\sqrt{2}} ,
 \theta_L=\theta_R-\theta_l , \tan\theta_l=\frac{M_{U_3}}{m_R+m_L}$ and $\tan 2\theta_R=\frac{2 M_{U_3}  m_R}{m_R^2-m_L^2-M_{U_3}^2}. 
$
\section{Z FCNC for top quark and its partner }
One can compute the Z FCNC  by rewriting left and right up type qaurks in SU(2) doublet in terms of top quark $t$ and its partner $t^\prime$,
\bea
u_{3R}&=&-\sin \theta_R t_R + \cos \theta_R t_R^\prime, \nn  \\
u_{3L}&=&\cos \theta_L t_L + \sin \theta_L t_L^\prime. 
\label{eq:u3}
\eea
With Eq.(\ref{eq:u3}) and Eq.(\ref{eq:I3}), Z FCNC for top quark and top prime quark is,
\bea
&& {\cal L}_{\rm Z \ I_3}^{t t^\prime}=
 -\frac{g_R \sin \theta_{13} Z^\mu }{2 \cos \theta_{W_R}} \left\lbrace  \sin^2 \theta_R \overline{t_R} \gamma_\mu t_R
-\sin \theta_R \cos \theta_R  \left( \overline{t_R} \gamma_\mu t_R^\prime + h.c.\right) 
+\cos^2 \theta_R \overline{t^\prime_R} \gamma_\mu t^\prime_R \right\rbrace 
\nn \\
&&  -\frac{Z^\mu( g_L \cos \theta_{13} + e \tan \theta_{W_R}  \sin \theta_{13} )} {2 \cos \theta_W}  \left\lbrace 
\cos^2 \theta_L \overline{t_L} \gamma_\mu t_L
+\sin \theta_L \cos \theta_L  \left( \overline{t_L} \gamma_\mu t_L^\prime + h.c.\right) 
+\sin^2 \theta_L \overline{t^\prime_L} \gamma_\mu t^\prime_L 
\right\rbrace .   \nn \\
\eea
When $M_{U_3} \ll  m_R$, one can show $\theta_L\simeq 0 \ll \theta_R \simeq \frac{M_{U_3}}{m_R} \ll 1$ .
\section*{Acknowledgment}
We would like to thank organizers of  KAON 2022. The work of T. M. is supported by Japan Society for the Promotion of Science
(JSPS) KAKENHI Grant Number JP17K05418 and JP16H03993. 
%A.H.P would like to thank the Ministry of Education, Culture, Sports, Science, and %Technology, Japan (MEXT) for the financial support during this work.

\end{document}